\documentclass[a4paper]{jpconf}
\usepackage{array}
\usepackage{color}
\usepackage{graphicx}
\usepackage{amsmath,amssymb,amstext}
\usepackage{multirow}
\usepackage{rotating}
\usepackage{float}
\usepackage{mathptmx}
\usepackage[numbers]{natbib}
\usepackage{soul,xcolor} 
\setul{-.6ex}{.3ex} 
\setstcolor{red}
\def\bra#1{\bigl\langle{ #1} \bigr|}
\def\ket#1{\bigl|{ #1} \bigr\rangle}

\def\ovlp#1#2{\bigl\langle{ #1}\big|{#2} \bigr\rangle}

\def\rvec {{\bf r}}

\def\qvec {{\bf q}}

\def\kvec {{\bf k}}

\def\creat#1{a_{#1}^{\dagger}}
\def\annil#1{a_{#1}^{\phantom{\dagger}}}            
\def\he#1{$^{#1}$He}
\def\EF{e_{\rm F}}
\def\KF{k_{\rm F}}
\def\SF{S_{\rm F}}
\def\a0{a_0}
\def\I{{\rm i}}

\begin{document}
\title{An analysis of variational wave function for the pairing
  problem in strongly correlated system}

\author{H.-H. Fan$^{\dagger\ddagger}$ and
  E.~Krotscheck$^{\dagger\ddagger}$}

\address{$\dagger$Department of Physics, University at Buffalo, SUNY
Buffalo NY 14260}
\address{$^\ddagger$Institut f\"ur Theoretische Physik, Johannes
Kepler Universit\"at, A 4040 Linz, Austria}

\ead{eckhardk@buffalo.edu}

\begin{abstract}

  We report a theoretical analysis of variational wave functions
  for the BCS pairing problem.  Starting with a Jastrow--Feenberg (or, in
  a more recent language ``fixed--node'') wave function for the
  superfluid state, we develop the full optimized Fermi-Hypernetted
  Chain (FHNC-EL) equations which sum a local approximation of the
  parquet--diagrams. Close examination of the procedure reveals that
  it is essential to go {\em beyond\/} the usual Jastrow--Feenberg
  approximation to guarantee the correct stability range.

\end{abstract}

\section{Motivation}
\label{sec:intro}

Jastrow--Feenberg theory has been established as a
very effective method for calculating ground state properties
of strongly interacting systems. We assume a
non-relativistic many-body Hamiltonian
\begin{equation}
H = -\sum_{i}\frac{\hbar^2}{2m}\nabla_i^2 + \sum_{i<j}
V(i,j)\,,
\label{eq:Hamiltonian}
\end{equation}
or, in second quantized form,
\begin{equation}
\hat H = \sum_{\alpha,\beta}\langle \alpha |T|\beta\rangle \creat{\alpha}\annil{\beta} +\frac{1}{2} \sum_{\alpha,\beta,\gamma,\delta}\langle \alpha \beta|V|\gamma \delta\rangle \creat{\alpha}\creat{\beta}\annil{\delta}\annil{\gamma}\,.
\label{eq:Hamiltonian2}
\end{equation}

The method starts with an {\em ansatz\/} for the wave function,
\cite{FeenbergBook}
\begin{eqnarray}
\Psi_0({\bf r}_1,\ldots,{\bf r}_N) &=& \frac{1}{\sqrt{I_{{\bf o},{\bf o}}}}
        F_N({\bf r}_1,\ldots,{\bf r}_N)
        \Phi_0({\bf r}_1,\ldots,{\bf r}_N)\label{eq:wavefunction},\\
F_N({\bf r}_1,\ldots,{\bf r}_N) &=& \exp\frac{1}{2}
\left[\sum_{i<j}  u_2({\bf r}_i,{\bf r}_j) + \cdot\cdot +
\sum_{i_1<\ldots<i_n}u_n({\bf r}_{i_1},.., {\bf r}_{i_n})
+ \cdot\cdot \right]\,,
\label{eq:Jastrow}
\end{eqnarray}
where ${I_{{\bf o},{\bf o}}} = \left\langle \Phi_0 | F_N^\dagger F_N |
\Phi_0\right\rangle$ is the normalization constant.  Here $\Phi_0({\bf
  r}_1,\ldots,{\bf r}_N)$ denotes a model state, which for normal
Fermi systems is a Slater-de\-ter\-mi\-nant, and $F_N$ is an $N$-body
correlation operator written in general form, but to be truncated at
the two-body $u_2$ term in a standard Jastrow calculation. The
computationally most effective way to deal with expectation values of
correlated wave functions of the type (\ref{eq:Jastrow}) are
diagrammatic methods, specifically the optimized Euler-Lagrange
Fermi-hypernetted chain (FHNC-EL) method, which is well suited for
calculation of physically interesting quantities.  These diagrammatic
methods have been successfully applied to such highly correlated Fermi
systems as $^3$He at $T=0$~\cite{polish}. We have shown in recent work
\cite{ljium} that even the simplest version of the FHNC-EL theory is
accurate within better than one percent at densities less than 25\% of
the saturation density of liquid \he3 and similarly for nuclear
systems \cite{ectpaper}.  It is one of the most attractive features of
the FHNC-EL method that the Euler equation (See Eq. (\ref{eq:euler})
below) has no solution if the assumed ground state is incorrect.

Following up on previous work \cite{HNCBCS,CBFPairing} and partly
motivated by interest in the BCS-BEC crossover in cold gases (see
Refs. \citenum{duinePhysRep04} and \citenum{chenPhysRep05} for review
articles), we have recently examined pairing phenomena in both model
Fermi systems \cite{cbcs} and neutron matter interacting via the
$^1S_0$ components of the Reid soft-core $V_6$ \cite{Reid68} and the
Argonne $V_{4}'$ \cite{AV18} two-nucleon interactions. These
calculations were based on a generalization of the BCS wave function

\begin{equation}
\ket{\rm BCS} =
{\prod_{\kvec}}
\left[ u_{\kvec} +  v_{\kvec} a_{ {\bf  k} \uparrow }^\dagger
 a_{-{\bf  k} \downarrow}^\dagger  \right] |0 \rangle\,
\label{eq:BCS}
\end{equation}
of the form
\begin{equation}
\ket{\rm CBCS} =  \sum_{{\bf m},N} \ket {\Psi_{\bf m}^{(N)}}
\ovlp{{\bf m}^{(N)}}{\rm BCS}
\label{eq:CBCS}
\end{equation}
where the correlated states $N$-body $\ket {\Psi_{\bf m}^{(N)}}$ are
\begin{eqnarray}
  \ket{ \Psi_{\bf m}^{(N)} } &=& \left[I_{{\bf m},{\bf m}}^{(N)}\right]^{-1/2}F_N\ket{{\bf m}^{(N)}}\,,
  \qquad I_{{\bf m},{\bf m}}^{(N)}\equiv 
{\bra{{\bf m}^{(N)} } F_{\!N}^{\dagger} F^{\phantom{\dagger}}_{\!N} \ket{{\bf m}^{(N)} }}
 \,,\label{eq:Psim}
\end{eqnarray}
Here the $\{\ket{{\bf m}^{(N)}}\}$ form complete sets of $N$-particle
Slater determinants of single-particle orbitals. We have commented on
alternative choices of the correlated BCS wave function in
Ref. \citenum{cbcs}, the arguments do not need to be repeated here.

Physically interesting quantities like the free energy of the
superfluid system
\begin{equation}
  \left\langle H'\right\rangle_s = \frac{\bra{\mathrm{CBCS}} \hat H'
    \ket{\mathrm{CBCS}}}
  {\bigl\langle\mathrm{CBCS}\big|\mathrm{CBCS}\bigr\rangle}\,,\qquad
  \hat H' \equiv \hat H - \mu\hat N\,.
  \label{eq:EBCS}
\end{equation}
are then calculated by cluster expansion and resummation techniques;
the correlation functions are determined by the variational principle
\begin{equation}
\frac{\delta  \left\langle H'\right\rangle_s }
{\delta u_n}({\bf r}_1,\ldots,{\bf r}_n) = 0\,.
\label{eq:euler}
\end{equation}

The combination of the (Fermi-)hypernetted-chain summation technique
for the energy with the optimization of the ground state correlations
is equivalent to a local approximation of the so--called
``parquet-diagrams'' \cite{parquet1,parquet2,parquet3}.

In Refs. \citenum{cbcs} and \citenum{ectpaper} we have simplified the
problem by expanding the free energy (\ref{eq:EBCS}) in the {\em
  deviation\/} of the Bogoliubov amplitudes $u_{\kvec}$, $v_{\kvec}$
from their normal state values $u^{(0)}_{\kvec}= \theta(k-\KF)$,
$v^{(0)}_{\kvec}=\theta(\KF-k)$. In that case, the pair correlation
functions $u_2(r)$ can be optimized for the normal system. With that,
we have arrived at the energy expression of the superfluid state
\begin{eqnarray}
\langle \hat H' \rangle_s &=& H_{00}^{(N)} - \mu N + 2 \sum_{\kvec
  ,\,|\, \kvec \,|\,>\KF} v_{\kvec}^2 (e_{\kvec} - \mu ) - 2
\sum_{\kvec , \,|\, \kvec \,|\,<\KF} u_{\kvec}^2 (e_{\kvec} - \mu )
\nonumber \\ &\quad& + \sum_{\kvec,\kvec'}u_\kvec v_\kvec u_{\kvec'}
v_{\kvec'} {\cal P}_{\kvec\kvec'}\,.
\label{5.7.13}
\end{eqnarray}
Above, $H_{00}^{(N)}$ is the energy of the normal $N$-particle system,
$\mu$ is the chemical potential. The $e_{\kvec}$ are the single
particle energies derived in correlated basis function (CBF) theory
\cite{CBF2}, and the paring interaction has the form
\begin{eqnarray}
{\cal P}_{\kvec\kvec'} &=& {\cal W}_{\kvec\kvec'}+(|e_{\kvec}- \mu | 
+ |e_{\kvec'}- \mu |)
{\cal N}_{\kvec\kvec'}\label{eq:Pdef}\,,\\
{\cal W}_{\kvec\kvec'} &=& \bra{\kvec \uparrow ,-\kvec\downarrow}
{\cal W}(1,2)\ket{\kvec'\uparrow ,-\kvec'\downarrow}_a\,,\label{eq:Wdef}\\
{\cal N}_{\kvec\kvec'}&=&
\bra{\kvec \uparrow ,-\kvec\downarrow}
{\cal N}(1,2)\ket{\kvec'\uparrow , - \kvec'\downarrow}_a\,.
\label{eq:Ndef}\end{eqnarray}
where the effective interaction ${\cal W}(1,2)$ and the correlation
corrections ${\cal N}(1,2)$ are then given by the
compound-diagrammatic ingredients of the FHNC-EL method for
off-diagonal quantities in CBF theory \cite{CBF2}.

The Bogoliubov amplitudes
$u_\kvec $, $v_\kvec $ are obtained in the standard way by variation
of the energy expectation (\ref{5.7.13}).  This leads to the familiar
gap equation
\begin{equation}
\Delta_\kvec = -\frac{1}{2}\sum_{\kvec'} {\cal P}_{\kvec\kvec'}
\frac{\Delta_{\kvec'}}{\sqrt{(e_{\kvec'}-\mu)^2 + \Delta_{\kvec'}^2}}\,.
\label{eq:gap}
\end{equation}
In what follows, we shall denote the value of the gap function
$\Delta_\kvec$ at the Fermi surface with
$\Delta\equiv\Delta_{|\kvec|=\KF}$.

The conventional ({\em i.e.\/} ``uncorrelated'' or ``mean-field'') BCS
gap equation \cite{FetterWalecka} is retrieved by replacing the
effective interaction ${\cal P}_{\kvec\kvec'}$ by the pairing matrix
of the bare interaction.

In the calculations of Refs. \citenum{ectpaper} and \citenum{cbcs} we have,
however, encountered two problems indicating that the  ``weak
coupling approximation'' can be problematic:
\begin{itemize}
\item{} We have examined in Ref. \citenum{cbcs} a large array of model
  systems described by a Lennard-Jones potential
  \begin{equation}
V_{\rm LJ} = 4\epsilon \left[\left(\frac{\sigma}{r}\right)^{12}
-\left(\frac{\sigma}{r}\right)^{6}\right]
\label{eq:VLJ}
\end{equation}
and an attractive square well (SW) potential
\begin{equation}
V_{\rm SW}(r) =
\begin{cases}
-\epsilon & \text{if}\quad r < \sigma\,, \\
\phantom{-} 0&  \text{if}\quad r > \sigma\,.\\
\end{cases}
\end{equation}

Both potentials are parametrized by a characteristic length $\sigma$
and the depth $\epsilon$ of the attractive well.  For a large array of
model systems with net attractive interactions (measured by a negative
vacuum scattering length $a_0$) we found that the {\em in--medium
  scattering length\/} $a$ diverges, as a function of interaction
strength, well before the vacuum scattering length $a_0$ diverges,
as shown in Fig. \ref{fig:amed_all}.
\begin{figure}
\begin{minipage}{0.99\textwidth}
\hspace{0.1cm}
\begin{minipage}{0.48\textwidth}
\centering
\includegraphics[width=8cm]{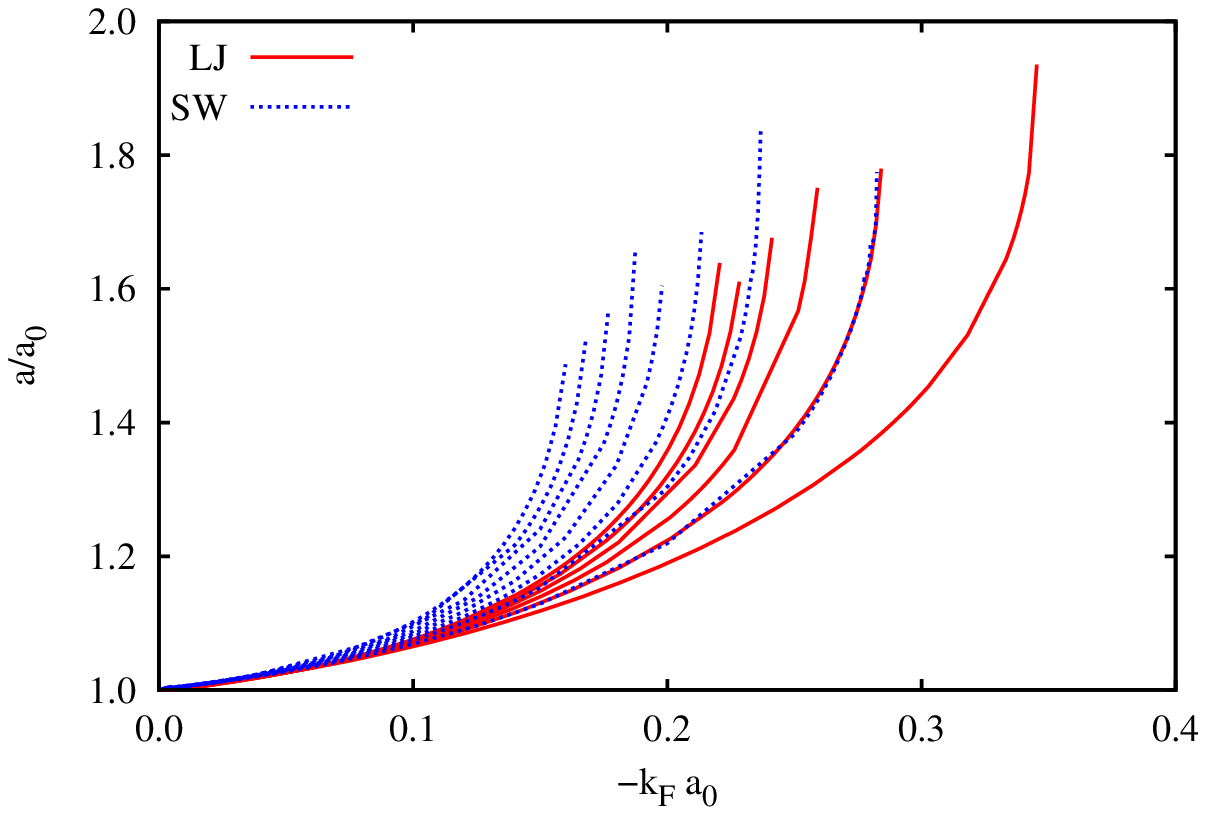}
\end{minipage}
\hspace{0.2cm}
\begin{minipage}{0.48\textwidth}
\centering \vspace{0cm}
\centerline{\includegraphics[width=5.4cm, height=8.2cm ,angle=-90]{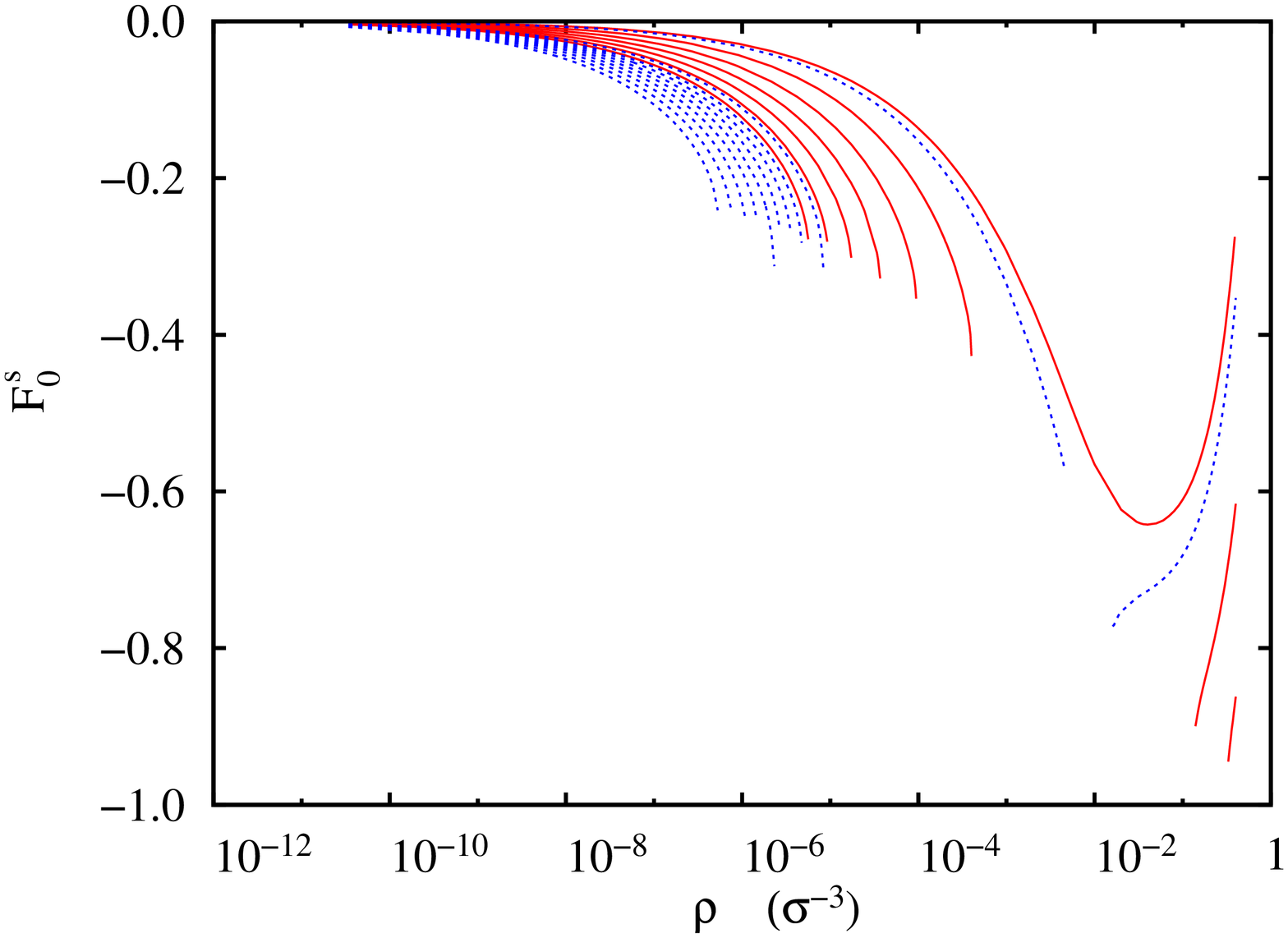}}
\end{minipage}
\end{minipage}

\caption{The left plot shows the ratio between the in-medium
  scattering length $a$ and the vacuum scattering length $\a0$ as
  function of $-\KF\a0$, for the Lennard-Jones (red lines) and the
  square-well potential (blue lines). The curves correspond to values
  of $\a0/\sigma=-0.5,-1.0,\dots,-4.0$ (SW) and
  $\a0/\sigma=-1.5,-2.0,\dots,-4.0$ (LJ), with the higher curves
  corresponding to larger $|\a0|$.  The right plot shows the
  dependence of the Landau parameter $F_0^s$ for the Lennard-Jones
  model of the interaction. The blue (dashed) lines correspond to
  coupling strengths $\epsilon = 7.0, 9.0, 9.2, 9.3,\ldots 9.9$.  The
  curves for the interaction strengths $\epsilon =
  7.0\ (\a0/\sigma=-1.12)$ (blue, dashed), $\epsilon =
  7.51\ (\a0/\sigma=-1.5)$ and $\epsilon = 8.01\ (\a0/\sigma=-2)$
  (both red, solid) are discontinuous at high density. The curve that
  ends at the lowest density corresponds to the strongest
  interaction. From Ref. \citenum{cbcs}.}
\label{fig:amed_all}
\end{figure}

The divergence of the in-medium scattering length appears to be an
indication for the formation of bound dimers (in other words the
BCS-BEC ``crossover'') which is enhanced by phonon exchange. Such a
dimerized state is not described by the Jastrow--Feenberg wave
function of a normal system. On the other hand, the simple BCS wave
function (\ref{eq:BCS}) can describe the crossover
\cite{CMS,NozieresSchmittRink} in situations where many--body
correlations are negligible. It is therefore expected that the
correlated BCS state (\ref{eq:CBCS}) can describe the system through
the BCS-BEC crossover, taking many--body correlations like
phonon--exchange properly into account.

\item{} For systems like neutron matter that have no many-body bound
  state, many--body correlations are less important. Nevertheless we
  found in Ref. \citenum{ectpaper} -- in agreement with very many
  previous calculations and also more recent fixed--node Monte-Carlo
  calculations \cite{GC2008,GC2010} at low densities, a pairing gap of
  the order of half of the Fermi energy $\EF$, see
  Fig. \ref{fig:gapef}. The derivations of the correlated BCS theory
  \cite{cbcs,HNCBCS,CBFPairing,KroTrieste} assume, on the other hand,
  that the paring gap is much smaller than the Fermi energy; the
  validity of the approximations leading to the formulation
  (\ref{5.7.13}-\ref{eq:Ndef}) must therefore be questioned.

\begin{figure}
\begin{minipage}{0.99\textwidth}
\vspace{-0.1cm}
\begin{minipage}{0.55\textwidth}
\centerline{\includegraphics[height=8.2cm, angle=-90]{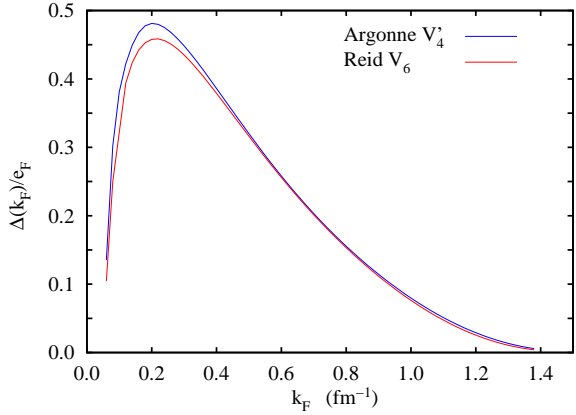}}
\end{minipage}
\hspace{0.5cm}
\begin{minipage}{0.40\textwidth}
  \caption{The figure shows the ratio $\Delta/\EF$ in neutron matter
    for the Reid $V_6$ and the Argonne $V_4'$ potentials. From
    Ref. \citenum{ectpaper}.\label{fig:gapef} }
\end{minipage}
\end{minipage}
\end{figure}
\end{itemize}

\section{BCS theory with local correlations}
\label{eq:cbcs}
 The divergence of the Euler equation observed in Ref. \citenum{cbcs}
 -- recall that we have assumed there a normal ground state and
 treated BCS correlations perturbatively -- is therefore expected to
 be a signature of a crossover to a strongly dimerized state: The
 disappearance of solutions of the Euler equation for the {\em
   normal\/} system tells us that our assumption that the system is
 normal is incorrect, and the pairing is so strong that a perturbative
 expansion in terms of the deviation from the normal state as used
 previously is not legitimate.

To clarify the situation, we have developed the
cluster expansion and resummation procedures for the fully correlated
state (\ref{eq:CBCS}).

The central quantity in the development of the method is the free
energy (\ref{eq:EBCS}) of the superfluid system which we can write
as
\begin{equation}
  \left\langle H'\right\rangle_s
  =  \frac{\sum_{N,{\bf m},{\bf n}} \ovlp{\rm BCS}{{\bf m}^{(N)}}
  \bra{\Psi_{\bf m}^{(N)}}\hat H'\ket{\Psi_{\bf n}^{(N)}}
\ovlp{{\bf n}^{(N)}}{\rm BCS}}{\ovlp{\rm CBCS}{\rm CBCS}}\,.
\label{eq:EBCSsum}
\end{equation}
It is helpful in the development of the formal theory to utilize
methods developed for the cluster expansions of the normal system
\cite{CBF2}. Besides the normalization integrals $I_{{\bf m},{\bf m}}$
defined in Eq. (\ref{eq:Psim}) we need the overlap integrals
\begin{equation}
    M_{\mathbf{m},\mathbf{n}}^{(N)} \equiv  \ovlp{\Psi_{\bf m}^{(N)}}{\Psi_{\bf n}^{(N)}}
   \equiv \delta_{\mathbf{m},\mathbf{n}} + N_{\mathbf{m},\mathbf{n}}^{(N)}\,.
   \label{eq:Nmndef}
\end{equation}
and matrix elements of the Hamiltonian
\begin{eqnarray}
  H_{\mathbf{m},\mathbf{n}}^{(N)} &\equiv& \bra{ \Psi_{\bf m}^{(N)} } \hat H'\ket{ \Psi_{\bf m}^{(N)} }\label{eq:Hmndef}\\
  W_{\mathbf{m},\mathbf{n}}^{(N)} &\equiv&H_{\mathbf{m},\mathbf{n}}^{(N)}
  - \frac{1}{2}
  \left(H_{\mathbf{m},\mathbf{m}}^{(N)}+H_{\mathbf{n},\mathbf{n}}^{(N)}\right)
  M_{\mathbf{m},\mathbf{n}}^{(N)}\label{eq:Wmndef}\,.
\end{eqnarray}
Using these definitions, we can write the superfluid free energy as
\begin{eqnarray}
  \left\langle H'\right\rangle_s &=& E_{\rm diag} +
  \frac{\sum_{N,{\bf m},{\bf n}} \ovlp{\rm BCS}{{\bf m}^{(N)}}W_{\mathbf{m},\mathbf{n}}^{(N)}\ovlp{{\bf n}^{(N)}}{\rm BCS}}{\ovlp{\rm CBCS}{\rm CBCS}}
  \nonumber\\
  &+&\frac{1}{2}\frac{\sum_{N,{\bf m}\ne{\bf n}}
    \ovlp{\rm BCS}{{\bf m}^{(N)}}\left(H_{\mathbf{m},\mathbf{m}}^{(N)}
    + H_{\mathbf{n},\mathbf{n}}^{(N)}-2E_{\rm diag}\right)
    N_{{\bf m},{\bf n}}\ovlp{{\bf n}^{(N)}}{\rm BCS}}{\ovlp{\rm CBCS}{\rm CBCS}}\nonumber\\
  &\equiv& E_{\rm diag} + E_{\rm offd} + E_{\rm enum}
\label{eq:FCBCS}
\end{eqnarray}
where
\begin{equation}
  E_{\rm diag} = \sum_{N,{\bf m}} \ovlp{\rm BCS}{{\bf m}^{(N)}}
  H_{\mathbf{m},\mathbf{m}}^{(N)}\ovlp{{\bf m}^{(N)}}{\rm BCS}\,.
\end{equation}

The actual expansions are tedious matter \cite{FanThesis} and will be
reported elsewhere, we display here only the most important results
and discuss, in a simplified example, the salient conclusions of our
analysis. The {\em physical interpretation\/} of the three terms in
Eq. (\ref{eq:FCBCS}) is revealed by examining cluster expansions for
the individual quantities. It turns out that the term $E_{\rm diag}$ is
obtained simply the expectation value of a state where the momentum
distribution of the ground state given by $v_{\kvec}^2$ instead of the
normal Fermi distribution $v^{(0)}_{\kvec}=\theta(\KF-k)$.  The second
term in Eq. (\ref{eq:FCBCS}) contains only off--diagonal matrix
elements, we can order this term according to the {\em number of
  orbitals\/} in which the states $\ket{{\bf m}^{(N)}}$ and $\ket{{\bf
    n}^{(N)}}$ differ, which will turn out, in the diagrammatic
analysis, the number of Cooper pairs. If pairing is weak as examined
in our previous work, then $\ket{{\bf m}^{(N)}}$ and $\ket{{\bf
    n}^{(N)}}$ differ by only two states. In other
words we consider the interaction of only one Cooper pair at a time.

The last term, $E_{\rm enum}$ gives rise to the ``energy numerator
corrections'' shown in (\ref{eq:Pdef}).  A discussion of the
significance of these terms is found in Refs.  \citenum{cbcs} and
\citenum{ectpaper}.

The purpose of the decomposition (\ref{eq:FCBCS}) is that cluster
expansion and resummation techniques of $(E_{\rm diag} + E_{\rm offd})$ can be derived
directly from the corresponding expression for the normal system. We
cite here only the final result:

In the {\em normal\/} system, the cluster expansion of the generating
function $G\equiv\ln {I_{{\bf o},{\bf o}}}$ is diagrammatically
expressed by the sum of all irreducible diagrams constructed by the
following rules \cite{Johnreview}:

\begin{enumerate}
\item{} Small circles depict particle coordinates. Filled circles
  imply integration over the particle coordinate and a density factor.
\item Every point is attached to the diagram by at least
one correlation line $h(r) \equiv \exp(u_2(r))-1$.
\item Every pair of points is connected by at most one correlon
line.
\item Exchange lines $\ell(r\KF)$ always appear in non-overlapping
  closed loops, or polygons. An $n-$sided exchange polygon contributes
  a weight factor $(-1/\nu)^{n-1}$ to the corresponding analytic
  expression, where $\nu$ is the degree of degeneracy of
  single--particle states.
\item No point can be attached to more than two
exchange lines.
\end{enumerate}

The corresponding expansion for the {\em superfluid\/} system is
generated from that by the following rule:
\begin{enumerate}
\item{} Intepret the density factor as
\begin{equation}
  \rho = \frac{\nu}{\Omega}\sum_{\kvec} v_{\kvec}^2\,,
\end{equation}
where $\Omega$ is the normalization volume.
  \item{} re-interpret all exchange lines $\ell(r\KF)$ as
\begin{equation}
    \ell_v(r) \equiv \frac{\nu}{\rho\Omega}
    \sum_{\kvec} v_{\kvec}^2e^{\I\kvec\cdot\rvec} = \frac{\nu}{\rho}
\int \frac{d^3 k}{(2\pi)^3} v_{\kvec}^2e^{\I\kvec\cdot\rvec}  \,.  
  \label{eq:lvdef}
\end{equation}
\item{} In each exchange loop, replace, in turn, each pair of exchange
  lines $\ell_v(r_{ij})\ell_v(r_{kl})$ by a pair
  $-\ell_u(r_{ij})\ell_u(r_{kl})$, where
    \begin{equation}
   \ell_u(r) \equiv \frac{\nu}{\rho\Omega}
   \sum_{\kvec} u_{\kvec}v_{\kvec}e^{\I\kvec\cdot\rvec} =  \frac{\nu}{\rho}
\int \frac{d^3 k}{(2\pi)^3} u_{\kvec}v_{\kvec}e^{\I\kvec\cdot\rvec}\,.
   \label{eq:ludef}
\end{equation}
\end{enumerate}

\begin{figure}[H]
\centerline{\includegraphics[width=0.8\columnwidth]{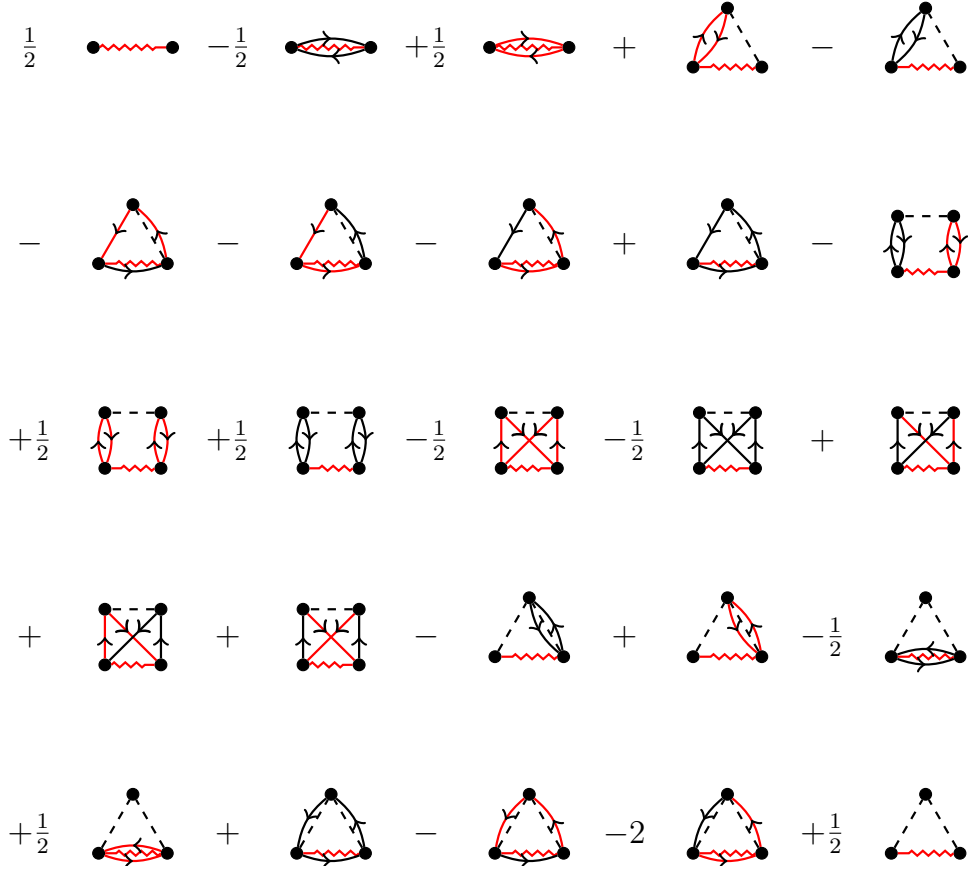}}
\caption{(color online) The figure shows the leading diagrams
  contributing to the potential energy. The black and red exchange
  lines depict the exchange functions $\ell_v(r_{ij})$ and
  $\ell_u(r_{ij})$, {\em cf.}  Eqs.  (\ref{eq:lvdef}) and
  (\ref{eq:ludef}), respectively, and the red wavy line an interaction
  $V(r_{ij})$. Otherwise we follow the usual diagrammatic conventions
  \cite{Johnreview}.  The uniform limit approximation amounts to
  keeping only diagrams 1-5, 10-12, and 25. The term $E_{\rm diag}$ is
  represented by the subset of diagrams containing only black
  exhange lines.
\label{fig:epot}}
\end{figure}

The FHNC theory for the terms $E_{\rm diag}+E_{\rm offd}$ in the
energy expression (\ref{eq:FCBCS}) can therefore be read off from the
diagrammatic expansions for the normal system. As an example, we show
in Fig. \ref{fig:epot} the diagrammatic representation of some leading
potential energy contributions to $E_{\rm diag}+E_{\rm offd}$.

The energy numerator terms must be treated separately, we will spell
out in the next section the simplest possible form.

\subsection{Uniform limit approximation}

We shall not implement the full method here but rather spell out what
is known as the ``uniform limit'' approximation because it shows
already the most important features of the theory and implications of
using locally correlated wave functions in the superfluid
system. Technically, the uniform limit approximation amounts to
assuming that a product of any two correlation functions is negligible
when it appears in coordinate space, but not negligible in momentum
space, in other words we assume $h^2(r) \ll h(r)$ but not that $\tilde
h^2(k)$ is small compared to $\tilde h(k)$.  Note that we define
here the dimensionless Fourier transform
\begin{equation}
  \tilde f(k) = \rho\int d^3r e^{\I\rvec\cdot\kvec}f(r)\label{eq:ft}\,.
\end{equation}

In the uniform limit approximation, only those exchange diagrams are
retained that contain either a loop $\ell_v^2(r_{ij})$ or
$\ell_u^2(r_{ij})$, see Fig. \ref{fig:epot}.  It is then convenient
to define
\begin{eqnarray}
  \sigma_v(r) &=& \frac{1}{\rho}\delta(r)-\frac{1}{\nu}\ell_v^2(r)
  \label{eq:sigmavdef}\\
  \sigma_u(r) &=& \frac{1}{\nu}\ell_u^2(r)
  \label{eq:sigmaudef}\,.\nonumber\\
\end{eqnarray}

Then, the static structure function of the ``non--interactig'' system
is
\begin{eqnarray}
  \SF(k) &=& \frac{1}{\left\langle N \right\rangle_0}
  \bra{\rm BCS}\rho_\kvec\rho_{-\kvec}\ket{\rm BCS}
 =\tilde{\sigma}_u(k) +\tilde{\sigma}_v(k) 
\,\label{eq:m0b}\\
  &=& 1 - \frac{1}{\nu}\left[\ell_v^2(r)\right]^{\cal F}(k)
  + \frac{1}{\nu}\left[\ell_u^2(r)\right]^{\cal F}(k)\label{eq:sofk}
\end{eqnarray}
where $[\ldots]^{\cal F}(k)$ is used as an alternative notation for
the Fourier--transform (\ref{eq:ft}) and $\left\langle N
\right\rangle_0 = \nu \sum_\kvec v_\kvec^2$ is the expectation value of
the number operator $\hat N$ with
respect to the BCS state (\ref{eq:BCS}).
\begin{eqnarray}
  \frac{E_{\rm diag}+E_{\rm offd}}{N}
  &=& \frac{T_{\rm BCS}}{N} + \frac{1}{2}\tilde V(0+) + \frac{1}{2}\int \frac{d^3k}{(2\pi)^3\rho}
  \left[S(k)-1\right]\tilde V(k)\nonumber\\
  &+&\frac{1}{2}\int \frac{d^3k}{(2\pi)^3\rho} \frac{t(k)}{2}\frac{(S(k)-\SF(k))^2}{\SF^2(k)S(k)}\,,
\label{eq:EpairsUL}
\end{eqnarray}
where $t(k) = \frac{\hbar^2k^2}{2m}$ and $T_{\rm BCS} =
\nu\sum_{\kvec} t(k) v_{\kvec}^2$ is the kinetic energy associated
with the BCS state (\ref{eq:BCS}).  In this approximation, the static
structure function is given by the chain approximation,
\begin{equation}
  S(k)
  = \frac{\SF(k)}{1-\tilde h(k)\SF(k)}\label{eq:SULofk}\,.
\end{equation}
We can, in this case, simply copy the equations from previous work,
with the understanding that the static structure function of the
non--interacting system, $\SF(k)$, is replaced by (\ref{eq:sofk}).

The evaluation of the energy numerator expression $E_{\rm enum}$ must
be done in an independent cluster expansion. These terms lead, in the
weakly coupled limit considered in previous work, to the second term
in the pairing matrix element (\ref{eq:Pdef}). Recall, however, the
derivation of the pairing matrix elements in Ref. \citenum{CBF2} and
the analysis of Ref. \citenum{cbcs} that the matrix elements
${\cal W}_{\kvec,\kvec'}$ should be identified with a static approximation
of the 2-body scattering matrix which contains ring-- and ladder--diagrams.
 In the uniform limit approximation, the pairing
interaction, on the other hand, does not contain ladder diagrams, and
the energy numerator term cancels the overcounted ladder diagrams in
the scattering matrix. The analysis is analogous to the one found in
Ref. \citenum{PethickSmith} where the gap equation is expressed in
terms of the $T$-matrix.

Now the ``uniform limit'' approximation contains only chain diagrams,
hence there is no double-counting of ladder diagrams and the energy
numerator term should, in that approximation, not be retained for the
derivation of the Euler equation. We have therefore now two
optimization conditions, one for the Bogoliubov amplitudes and one for
the pair correlations. The gap equation (\ref{eq:gap}) remains the
same, only the pairing matrix element depends implicitly on the
Bogoliubov amplitudes. The analysis of the Euler equation for
the spatial amplitudes is more complicated and will be carried
out in the next section.

\section{Euler equation for local correlations}

The analysis of the Euler equation for the spatial correlations
turns out to be much more subtle: Carrying out the variation
of the energy (\ref{eq:EpairsUL}) yields the familiar answer
\begin{equation}
  S(k) = \frac{\SF(k)}{\sqrt{1 + \displaystyle\frac{2\SF^2(k)}{t(k)}\tilde
      V(k)}}\label{eq:collRPA}\,.
\end{equation}
(Retaining the energy numerator terms simply supplements the bare
interaction with a small additive term). At the first glance,
the result looks innocuous. In the {\em normal\/} system, we have
$\SF(k) \propto k $ as $k\rightarrow 0+$ and the condition
that the expression under the square root be positive
\begin{equation}
1 + \frac{3}{4}\frac{3 m}{\hbar^2 \KF ^2} \tilde V(0+) > 0.
\label{eq:mc2coll}
\end{equation}
which is --- apart from the numerical factor 3/4 which will be
discussed momentarily --- recognized as the condition $F_0^s > -1$.
In the {\em superflud\/} system, we have instead
\begin{equation}
  \SF (0+) = 2\frac{\sum_{\kvec} u_{\kvec}^2 v_{\kvec}^2}{\sum_{\kvec}v_{\kvec}^2} > 0\,.
  \end{equation}

This says that a locally correlated wave function for a superfluid
system has a spurious instability whenever $\tilde V(0+)<0$, no matter
how small the gap is. This is evidently only a consequence of the
approximations implicit to the wave function (\ref{eq:CBCS}). We have
shown this here only for the case of a weakly interacting system, but
it is also true in the general case that all FHNC diagrams, and
possibly also higher--order correlation functions, are included: The
Euler equation remains structurally the same, only the bare potential
is replaced by an effective interaction \cite{polish}. Our observation
applies, of course, equally to ``fixed--node'' Monte Carlo
calculations which may see this instability only in large stochastic
fluctuations.  In view of the fact that the BCS-BEC crossover is
driven by a net attractive interaction (see also the right panel of
Fig. \ref{fig:amed_all}), it is worth examining this problem in
detail.

The source of the problem can be identified as follows:
In the {\em normal\/} system, the expression
(\ref{eq:collRPA}) can be derived from an ordinary random phase approximation
  \begin{equation}
    S(k) = -\int_0^\infty \frac{d\omega}{\pi} \Im m \frac{\chi_0(k,\omega)}
    {1-\tilde V(q)\chi_0(k,\omega)}\label{eq:SRPA}
  \end{equation}
where $\chi_0(k,\omega)$ is the Lindhard function.  Consistent with
the convention (\ref{eq:ft}) according to which $\tilde V(q)$ has the
dimension of an energy, we have defined the density-density response
function slightly different than usual \cite{FetterWalecka} namely
such that has the dimension of an inverse energy. Eq. (\ref{eq:collRPA})
can be obtained by approximating the Lindhard function 
$\chi_0(q,\omega)$ by a ``collective'' Lindhard function (occasionally
also referred to as ``one--pole approximation'' or ``mean spherical
approximation'') $ \chi_0^{\rm coll}(q,\omega)$ which is constructed by
approximating the particle--hole band by an effective single pole such
that the $m_0$ and $m_1$ sum rules are satisfied \cite{polish}
\begin{eqnarray}
        -\Im m\int_0^\infty\frac{ d\omega}{\pi}\chi_0^{\rm coll}(k,\omega) &&=
        -\Im m\int_0^\infty\frac{ d\omega}{\pi}\chi_0(k,\omega) = \SF(k)\label{m0MSA}\\
        -\Im m\int_0^\infty\frac{ d\omega}{\pi}\omega\chi_0^{\rm coll}(k,\omega) &&=
        -\Im m\int_0^\infty\frac{ d\omega}{\pi}\omega\chi_0(k,\omega) = t(k)\,.
\label{eq:m1MSA}
\end{eqnarray}
This leads to
\begin{equation}
        \chi_0^{\rm coll}(k,\omega) =
        {\displaystyle {2 t(k)}
        \over{ (\omega+\I\eta)^2-
        \left(\displaystyle{\frac{t(k)}{S_{\rm F}(k)}}\right)^2}} \,.
\label{eq:msa0}
\end{equation}
The frequency integration (\ref{eq:SRPA}) then can be carried out
analytically and leads to the equation (\ref{eq:collRPA}).  Experience
with electronic systems and the very strongly interaction \he3 has
shown that the expression (\ref{eq:collRPA}) is accurate within 1-2
percent; this is the reason why the Jastrow--Feenberg theory for
fermions has been so successful. Of course, this is relevant only for
integrated or Fermi--sea averaged quantities; we have pointed out
already a long time ago \cite{shores} that the approximation should be
particularly poor for observables that are determined by the dynamics
close to the Fermi surface.  Among others, the stability condition
(\ref{eq:mc2coll}) is replaced by
\begin{equation}
1 + \frac{3 m}{\hbar^2 \KF ^2} \tilde V(0+) > 0.
\label{eq:mc2full}
\end{equation}
which is the correct stability condition.

Returning to the superfluid case, one can expect that the correct
Lindhard function removes the spurious instability. There have been
several suggestions
\cite{PhysRevB.61.9095,PhysRevA.74.042717,Steiner2009,Vitali2017}.
The most frequently used form for $T=0$ is, in terms of the usual
relationships of BCS theory,
\begin{eqnarray}
  u_k^2 &=& \frac{1}{2}\left(1+\frac{\xi_\kvec}{E_\kvec}\right)
  \nonumber\\
  v_k^2 &=& \frac{1}{2}\left(1-\frac{\xi_\kvec}{E_\kvec}\right)\,.
\end{eqnarray}
with $\xi_{\kvec} = t(k)-\mu$ and $E_{\kvec} =
\sqrt{\xi_{\kvec}^2+\Delta_{\kvec}^2}$
we have \cite{Schrieffer1999,Kee1998,Kee1999,PhysRevB.61.9095}

\begin{equation}
  \chi_0(\qvec,\omega)
    = \frac{\nu}{\left\langle N \right\rangle_0}\sum_{\kvec}b_{\kvec,\qvec}^-\Biggl[\frac{1}
    {\omega-E_{\kvec+\qvec}-E_{\kvec}+\I\eta}
    - 
    \frac{1}{\omega+E_{\kvec+\qvec}+E_{\kvec}+\I\eta}\Biggr]
\end{equation}
where 
\begin{equation}
  b_{\kvec,\qvec}^- = \frac{1}{4}\left[1-
  \frac{\xi_{\kvec}}{E_{\kvec}}\frac{\xi_{\kvec+\qvec}}{E_{\kvec+\qvec}}
  +\frac{\Delta_{\kvec}}{E_{\kvec}}\frac{\Delta_{\kvec+\qvec}}{E_{\kvec+\qvec}}\right]\,.
\end{equation}
In the limit of a normal system, the coefficients $b_{\kvec,\qvec}^-$ become
\begin{equation}
  b_{\kvec,\qvec}^- \rightarrow  n_{\kvec}(1-n_{\kvec+\qvec})\,,
  \end{equation}
where $n_{\kvec}$ is the Fermi distribution, as it should come out.

This superfluid Lindhard function is consistent with the $\SF(k)$ as
defined in Eq. (\ref{eq:m0b}).

  \begin{eqnarray}
    \SF(k) &=& -\int_0^\infty \frac{d\omega}{\pi} \Im m \chi_0(k,\omega)
    = \frac{\nu}{\left\langle N \right\rangle_0}\int_0^\infty d\omega
    \sum_{\qvec}b_{\qvec,\kvec}^-\delta(\omega-E_{\qvec}-E_{\qvec+\kvec})\nonumber\\
    &=&  \frac{\nu}{\left\langle N \right\rangle_0}\sum_{\qvec}b_{\qvec,\kvec}^-
    =\tilde\sigma_u(k) + \tilde\sigma_v(k)
  \end{eqnarray}

  We can now return to the frequency integration (\ref{eq:SRPA}). All
  we need to show is that this expression exists, for small gaps, for
  $-1 < \frac{3m}{\hbar^2 \KF^2}\tilde V(0+)$.  For that, it is
  sufficient to look at the limit $q\rightarrow 0$ of the {\em
    static\/} response function and the {\em static\/} Lindhard
  function.

For $\omega=0$ we get for the static Lindhard function
\begin{equation}
  \lim_{q\rightarrow 0}\chi_0(q,0) = -\frac{\nu}{2\rho}\int\frac{ d^3k}{(2\pi)^3}
  \frac{\Delta_\kvec^2}{E_{\kvec}^3}\,.
\end{equation}

To estimate this limit for small gap energies, we add and subtract a
function that can be integrated exactly such that the remainder vanishes
as $\Delta\rightarrow 0$.
\begin{eqnarray}
  \lim_{q\rightarrow 0}
  \chi_0(q,0) &=& -\frac{4\pi\nu}{2\rho}\int  \frac{dk k^2}{(2\pi)^3}
  \frac{\Delta_\kvec^2}{\left((t(k)-\mu)^2+\Delta_\kvec^2\right)^{3/2}}
  \nonumber\\
  &=& -\frac{3}{2\KF^3}\int dk\Biggl[
    \frac{k^2 \Delta_\kvec^2}{\left((t(k)-\mu)^2+\Delta_\kvec^2\right)^{3/2}}
    - \frac{k\KF \Delta^2}{\left((t(k)-\mu)^2+\Delta^2\right)^{3/2}}
    \nonumber\\
    &&\qquad\qquad\qquad + \frac{k\KF \Delta^2}{\left((t(k)-\mu)^2+\Delta^2\right)^{3/2}}
    \Biggr]\nonumber\\
  &=& {\cal O}(\Delta)-\frac{3}{4}\left[
    \frac{1}{\sqrt{\mu^2+\Delta^2}}+\frac{1}{\mu}\right]
  \rightarrow -\frac{3}{2\mu}\quad\mathrm{as}\quad\Delta\rightarrow 0\,.
\end{eqnarray}
This is identical to the same limit of the Lindhard function of the
normal system, see Fetter-Walecka Eq. (12.46b), considering that the
Lindhard function defined here contains a factor $1/\rho$.

With that we get for the stability condition
\begin{equation}
  1 + \frac{3 \tilde V(0+)}{2\mu} > 0
\end{equation}

\section{Discussion}

We have formulated in this contribution the beginnings of a
theoretical method to deal with strongly interacting superfluids that
promises the same accuracy as what was achieved for the normal helium
liquids. We have studied the Euler equation for the local correlations
in a simplified example that does not require the formulation of the
full FHNC theory. We have derived the interesting, yet disturbing,
result that the Euler equation has no physically meaningful solution
for net attractive interactions $\tilde V(0+)<0$. We hasten to point
out that this result is not specific to the ``uniform limit
approximation'', in the fully correlated theory the bare interaction
$\tilde V(q)$ is simply replaced by the particle--hole interaction.
This is a rather profound observation since it applies to any local
correlation operator $F$; including the ``fixed--node'' approximation
used in Quantum Monte Carlo calculations where it would, of course, be
hard to discover.

We have also outlined a pathway to the solution of this problem: One
must go beyond local correlations. The {\em rigorous\/} way to do that
has been carried out in Ref. \citenum{rings}: One must add, to the
variational energy expectation value, an infinite sum of terms in CBF
perturbation theory.  That has the effect that the ``collective
approximation'' for $S(k)$, Eq. (\ref{eq:collRPA}) is replaced by the
RPA expression (\ref{eq:SRPA}). The calculations of
Ref. \citenum{rings} have been rather tedious; to be rigorous one
would have to re-do all of this work for the superfliud system. The
formulation of Coupled Cluster theory with correlated wave functions
\cite{CBFPairing} would be a way to carry this through, but, on the
other hand, our result is sufficiently plausible to be applicable
without a rigorous proof.

\ack

This work was supported, in part, by the College of Arts and
Sciences, University at Buffalo SUNY, and the Austrian Science Fund
project I602 (to EK). We would like to thank John Clark and Peter
Schuck for useful discussions on the subject of this paper.

\pagebreak


\begin{thebibliography}{10}
\expandafter\ifx\csname url\endcsname\relax
  \def\url#1{{\tt #1}}\fi
\expandafter\ifx\csname urlprefix\endcsname\relax\def\urlprefix{URL }\fi
\providecommand{\eprint}[2][]{\url{#2}}

\bibitem{FeenbergBook}
Feenberg E 1969 {\em Theory of {Q}uantum Fluids\/} (New York: Academic)

\bibitem{polish}
Krotscheck E 2000 {\em J. Low Temp. Phys.\/} {\bf 119} 103--145

\bibitem{ljium}
Egger J, Krotscheck E and Zillich R~E 2011 {\em J. Low Temp. Phys.\/} {\bf 165}
  275--291

\bibitem{ectpaper}
Fan H~H, Krotscheck E and Clark J~W 2017 {\em J. Low Temp. Phys.\/} {\bf 189}
  470--494

\bibitem{HNCBCS}
Krotscheck E and Clark J~W 1980 {\em Nucl. Phys. A\/} {\bf 333} 77--115

\bibitem{CBFPairing}
Krotscheck E, Smith R~A and Jackson A~D 1981 {\em Phys. Rev. B\/} {\bf 24}
  6404--6420

\bibitem{duinePhysRep04}
Duine R and Stoof H~T~C 2004 {\em Phys. Rep.\/} {\bf 396} 115--195

\bibitem{chenPhysRep05}
Chen Q, Stajic J, Tan S and Levin K 2005 {\em Phys. Rep.\/} {\bf 412}
  1--88

\bibitem{cbcs}
Fan H~H, Krotscheck E, Lichtenegger T, Mateo D and Zillich R~E 2015 {\em Phys.
  Rev. A\/} {\bf 92} 023640

\bibitem{Reid68}
{Reid, Jr} R~V 1968 {\em Ann. Phys. (NY)\/} {\bf 50} 411--448

\bibitem{AV18}
Wiringa R~B, Stoks V~G~J and Schiavilla R 1995 {\em Phys. Rev. C\/} {\bf 51}
  38--51

\bibitem{parquet1}
Jackson A~D, Lande A and Smith R~A 1982 {\em Phys. Rep.\/} {\bf 86}
  55--111

\bibitem{parquet2}
Jackson A~D, Lande A and Smith R~A 1985 {\em Phys. Rev. Lett.\/} {\bf 54}
  1469--1471

\bibitem{parquet3}
Krotscheck E, Smith R~A and Jackson A~D 1986 {\em Phys. Rev. A\/} {\bf 33}
  3535--3536

\bibitem{CBF2}
Krotscheck E and Clark J~W 1979 {\em Nucl. Phys. A\/} {\bf 328} 73--103

\bibitem{FetterWalecka}
Fetter A~L and Walecka J~D 1971 {\em {Q}uantum Theory of Many-Particle
  Systems\/} (New York: McGraw-Hill)

\bibitem{CMS}
Cooper L~N, Mills R~L and Sessler A~M 1959 {\em Phys. Rev.\/} {\bf 114}
  1377--1382

\bibitem{NozieresSchmittRink}
Nozi{\'e}res P and Schmitt-Rink S 1985 {\em J. Low Temp. Phys.\/} {\bf 59}
  195--211

\bibitem{GC2008}
Gezerlis A and Carlson J 2008 {\em Phys. Rev. C\/} {\bf 77} 032801

\bibitem{GC2010}
Gezerlis A and Carlson J 2010 {\em Phys. Rev. C\/} {\bf 81} 025803

\bibitem{KroTrieste}
Krotscheck E 2002 {\em Introduction to Modern Methods of {Q}uantum Many--Body
  Theory and their Applications\/} ({\em Advances in {Q}uantum Many--Body
  Theory\/} vol~7) ed Fabrocini A, Fantoni S and Krotscheck E (Singapore: World
  Scientific) pp 267--330

\bibitem{FanThesis}
Fan H~H 2018 {\em Pairing Phenomena from Low-Density Fermi Gases to Neutron
  Star Matter\/} Ph.D. thesis University at Buffalo SUNY

\bibitem{Johnreview}
Clark J~W 1979 {\em Progress in Particle and Nuclear Physics\/} vol~2 ed
  Wilkinson D~H (Oxford: Pergamon Press Ltd.) pp 89--199

\bibitem{PethickSmith}
Pethick C~J and Smith H 2008 {\em {B}ose-{E}instein Condensation in Dilute
  Gases\/} second edition ed (Cambridge, UK: Cambridge University Press)

\bibitem{shores}
Jackson A~D, Krotscheck E, Meltzer D and Smith R~A 1982 {\em Nucl. Phys. A\/}
  {\bf 386} 125--165

\bibitem{PhysRevB.61.9095}
Voo K~K, Wu W~C, Li J~X and Lee T~K 2000 {\em Phys. Rev. B\/} {\bf 13}
  9095--9100

\bibitem{PhysRevA.74.042717}
Combescot R, Kagan M~Y and Stringari S 2006 {\em Phys. Rev. A\/} {\bf 74}
  042717

\bibitem{Steiner2009}
Steiner A~W and Reddy S 2009 {\em Phys. Rev. C\/} {\bf 79} 015802 


\bibitem{Vitali2017}
Vitali E, Shi H, Qin M and Zhang S 2017 {\em J. Low Temp. Phys.\/} {\bf 189}
  312--327 


\bibitem{Schrieffer1999}
Schrieffer J~R 1999 {\em Theory Of Superconductivity (Advanced Books
  Classics)\/} revised ed (Perseus Books)

\bibitem{Kee1998}
Kee H~Y and Varma C~M 1998 {\em Phys. Rev. B\/} {\bf 58} 15035--15044


\bibitem{Kee1999}
Kee H~Y and Kim Y~B 1999 {\em Phys. Rev. B\/} {\bf 59} 4470--4474


\bibitem{rings}
Krotscheck E 1982 {\em Phys. Rev. A\/} {\bf 26} 3536--3556

\end{thebibliography}
\bibliographystyle{iopart-num}

\providecommand{\newblock}{}

\end{document}